\documentclass[useAMS,usenatbib]{mn2e}

\usepackage{amsmath,amssymb}
\usepackage{epsfig}

\newcommand{\HII}{H{\sc ii} }

\newcommand{\HeIII}{He{\sc iii} }

\newcommand{\simgt}{\lower.5ex\hbox{$\; \buildrel > \over \sim \;$}}

\title[21cm from early \HII regions]{The 21cm Signature of Early Relic \HII Regions}

\author[Tokutani et al]{Midori Tokutani$^{1}$
\thanks{E-mail:mtoku@a.phys.nagoya-u.ac.jp}, 
Naoki Yoshida$^{1,2}$, 
S. Peng Oh$^{3}$, 
Naoshi Sugiyama$^{1,2}$\\
$^{1}$Department of Physics, Nagoya University, Furocho, Nagoya, Aichi 464-8602, Japan\\
$^{2}$Institute for the Physics and Mathematics of the Universe, University of Tokyo, Kashiwanoha, Kashiwa, Chiba, Japan\\
$^{3}$Department of Physics, University of California, Santa Barbara, CA 93106, USA}

\begin{document}

\date{Submitted to MNRAS, September 1, 2008}

\pagerange{\pageref{firstpage}--\pageref{lastpage}} \pubyear{2008}

\maketitle

\label{firstpage}

\begin{abstract}

We calculate the spin temperature and 21 cm brightness of early
\HII regions around the first stars. We use outputs from cosmological
radiation-hydrodynamics simulations of the formation
and evolution of early \HII regions.
In the pre-reionization era, \HII regions around massive primordial stars 
have diameters of a few kpc. The gas within the \HII regions is almost fully
ionized, but begins recombining after the central stars die off. 
The relic \HII regions are then seen as bright {\it emission} sources in hydrogen 21 cm.
We make brightness temperature maps of the \HII regions, accounting for 
radiative coupling with Lyman-$\alpha$ photons
in a simplified manner. The spin temperature in the relic \HII region 
is close to the gas kinetic temperature, 
generally several hundred to several thousand degrees.
We show that the relic \HII region can be as bright as $\delta T_{\rm b} \sim 100$ mK
in differential temperature against the cosmic microwave background
for an angular resolution of sub-arcseconds.
While individual early \HII patches will not be identified by
currently planned radio telescopes, the collective fluctuations from early \HII regions
might imprint signatures in the 21 cm background.
\end{abstract}

\begin{keywords}
cosmology:theory - early universe - intergalactic medium
\end{keywords}

\section{Introduction}

Reionization of hydrogen in the intergalactic medium (IGM) is
an important landmark in the history of the universe. 
It is generally thought
that the sources of ionizing photons are early generations of stars 
and galaxies formed at $z > 6$.
Recent observations provided important constraints on the epoch of reionization;
the Gunn-Peterson troughs are found in the spectra of distant quasars
at $z = 6-7$ (White et al. 2003; Fan et al. 2006),
while the large-scale polarization of cosmic microwave background (CMB)
measured by the WMAP satellite suggests
that reionization 
may have begun at $z > 10$ (Page et al. 2007; Komatsu et al. 2008).

Observing 21 cm emission or absorption from hydrogen atoms in the high 
redshift IGM is a promising way of revealing the detailed process of reionization.
There have been a number of studies on 21 cm signature of large-scale
reionization by galaxies (McQuinn et al. 2006; Iliev et al. 2006;
see Furlanetto et al. 2006 for a review).
These studies are aimed at making forecasts for near-future 
radio observations such as LOFAR, and thus consider large-scale
structures of ionized/neutral IGM at $z \sim 6 - 10$.

There have been also theoretical studies on 21 cm signatures
from pre-reionization epochs.
Kuhlen et al (2006) study the 21 cm signature of early mini-quasars.
They show that X-rays from the mini-quasar raise the gas 
kinetic temperature and enhance 21 cm signals.
Chen \& Miralda-Escud$\acute{\rm e}$ (2004; 2008) argue that X-rays 
from the first stars heat the surrounding gas and couple the spin 
temperature to its kinetic temperature, generating a large Lyman-$\alpha$ absorption sphere.
Shapiro et al (2006) estimate the 21 cm fluctuations
caused by cosmological ``minihalos'' and by the IGM. 
They conclude that the 21 cm emission from minihalos dominates over
that from the diffuse IGM at $z \lesssim 20$.
Furlanetto \& Oh (2006) argue, however, that minihalos generate
only small 21 cm fluctuations. Clearly, it is important to identify
dominant sources of 21 cm 
fluctuations at high-redshifts.
None of these works, however, consider 21 cm signals from early {\it relic} \HII regions, 
which are hot and (partially) ionized even after the central sources' lifetimes,
and hence can be bigger and brighter than cosmological mini-halos.
Interestingly, in the standard theory of cosmic structure formation
based on cold dark matter, there is a large gap between the time
when the first stars are formed 
(Tegmark et al. 1997; Yoshida et al. 2003; 2008)
and when the IGM is {\it completely} ionized.
Hence it is expected that there were early {\it relic} \HII regions during the reionization process. 
Although individual \HII regions are too small to be detected even by future radio telescopes,
their collective signals may imprint distinguishable fluctuations.
In this paper, 
we study the 21 cm signature of 
early relic \HII regions around the first stars using cosmological
simulations. We assume that massive primordial stars are formed in early 
low-mass dark matter halos
and ionize a large volume of the surrounding IGM in their short lifetime of
a few million years. 
We compute the spin temperature and 21 cm differential temperature
fluctuations of the relic \HII regions.

\section{Numerical simulations}

We use the outputs of a cosmological simulation of
Yoshida et al. (2007, hereafter Y07). Briefly, they carried out 
three-dimensional radiation hydrodynamics calculations of the formation 
and evolution of early \HII / \HeIII regions around the first stars.  
Primordial gas chemistry such as hydrogen recombination is 
followed in a non-equilibrium manner in the simulation.
We generate the same realization as that of Y07 
but for a slightly lower value of the normalization
of the density fluctuation amplitude, $\sigma_{8} = 0.8$.
     
The other cosmological parameters are the same as in Y07,
$(\Omega_{\rm m},\, \Omega_{\rm b},\, \Omega_{\Lambda},\, H_{0})
= (0.26,\, 0.04,\, 0.7,\, 70[\rm{km}\! \cdot\! {\rm s}^{-1}\! \cdot\! {\rm Mpc}^{-1}])$,
which are consistent with the standard $\Lambda$CDM model (Komatsu et al. 2008).
Lowering the value of $\sigma_{8}$ simply shifts the
formation epoch of the first star to a later time. The main
properties of the \HII region are almost the same as those presented in Y07.
In our simulation, the first star-forming cloud is located at $z = 22$.
We embed a Population III star with a mass of 100$M_{\odot}$. 
In order to focus on the evolution of an early \HII region,
we assume that the central star dies without triggering supernova explosion.
Such a massive star may also leave behind a remnant black hole.
We defer detailed studies on the feedback effects from supernovae and from 
early remnant blackholes to future work. 
We note that early Population III remnant blackholes do not accrete the surrounding gas 
and emit X-rays efficiently, due to photoevaporation of gas from the host halo by the 
progenitor star, as well as photo-heating by the accreting black hole itself 
(Y07; Johnson \& Bromm 2007; Alvarez et al. 2008).
We further assume that UV radiation background
and UV radiation from other sources do not exist and 
that the central star is the first radiation source.
We follow the thermal evolution of the relic \HII region for over 
one hundred million years under these assumptions.

\section{21 cm signatures from early \HII regions}

The 21 cm line is produced via the transition 
between the triplet and the singlet sublevels of the hyperfine structure
of the ground level of a neutral hydrogen atom.
This wavelength corresponds to a frequency of 1420 MHz and a temperature of
$T_{\star} = 0.068 \rm K $.
The transition between the levels occurs through three processes:
absorption of CMB photons, collisions with other particles
(hydrogen atoms, free electrons and protons),
and scattering of UV photons.

The spin temperature is expressed as a weighted mean of these processes as
(Field 1958)
\begin{eqnarray}
T_{\rm S} =
\frac{ T_{\star} + T_{\rm CMB} + x_{\rm c}T_{\rm K} + x_{\alpha}T_{\rm c} }
{ 1 + x_{\rm c} + x_{\alpha}},
\label{eq:Tspin}
\end{eqnarray}
where $T_{\rm K}$ is the gas kinetic temperature,
$T_{\rm c}$ is the color temperature of Lyman-$\alpha$ photons,
and $x_{\rm c}$ and $x_{\alpha}$ are coefficients
for collisional and radiative coupling, respectively.

We do not consider a cosmological UV background radiation
because the 21 cm differential brightness 
temperature of relic \HII regions is not significantly affected 
by the strength of radiative coupling.
Also, we do not include the effect of heating by X-ray photons 
from the first stars. 
The gas outside the \HII region can be significantly 
heated by X-ray induced secondary electrons.
However, for a 100 $M_{\odot}$ star, 
the size of the heated region is only slightly larger than the extent 
of the \HII region (Chen \& Miralda-Escud$\acute{\rm e}$ 2008, Figures 3-5).
Hence ignoring the X-ray heating effect does not affect significantly
our results; in this case it would boost the observable signal by a 
factor of order unity. For very massive stars ($> 300 M_{\odot}$), 
which are hotter and generate more X-rays, the signal from the X-ray 
heated region could be significantly larger.

The coupling coefficient for collisions, $x_{\rm c}$, is given by
\begin{eqnarray}
x_{\rm c} = \frac{T_{\star}}{A_{10}T_{\rm K}}(C_{\rm H}+C_{\rm e}+C_{\rm p}),
\end{eqnarray}
where the spontaneous emission rate $A_{10}$ is $2.85 \times 10^{-15} {\rm s}^{-1}$,
and $C_{\rm H}$, $C_{\rm e}$ and $C_{\rm p}$ are the de-excitation 
rates for collisions with neutral atoms,
free electrons and protons, respectively
\footnote{Although the simulation of Y07 includes He atoms and ions, 
we ignore the collisions between H atoms and He species because 
the number fraction of He is small.}.
We use fitting functions of Zygelman (2005)
and Liszt (2001) for H-H collisions and e-H collisions, respectively.
For p-H collisions, we use the coefficient
$3.2$ times as large as that for H-H collisions (Smith 1966). 

The radiative coupling included in equation (\ref{eq:Tspin})
is known as the Wouthuysen-Field effect.
It is extremely difficult to evaluate accurately
the coupling coefficient for the incident UV radiation, $x_{\alpha}$,
because one needs to solve, in principle, full radiation transfer equations 
including the effect of absorption and re-emission of Lyman-$\alpha$ photons, 
the Hubble flow, and the injection of new photons.
By noting that tight radiative coupling is easily achieved under
many cosmological conditions (e.g., Furlanetto et al. 2006),
we assume that Lyman-$\alpha$ photons produced by recombination 
is sufficient for effective radiative coupling,
and examine the overall effect by setting a constant $x_{\alpha} = 0.5$ within \HII regions.
We compute the color temperature, $T_{\rm c}$, by solving the following equation
\begin{eqnarray}
T_{\rm c} = T_{\rm K}
\biggl( \frac{ 1 + T_{\rm se} / T_{\rm K} }
{ 1 + T_{\rm se} / T_{\rm S} } \biggl),
\label{eq:color}
\end{eqnarray}
where $T_{\rm se} = (2/9) T_{\rm K} {\nu}_{21}^2 / \Delta {\nu}_{ {\rm D},\alpha }^2$,
which is obtained by solving the radiative transfer equation for scattering
of UV photons for large optical depths (Furlanetto et al. 2006).
Here, ${\nu}_{21}$ is the frequency of 21 cm line,
$\Delta {\nu}_{ {\rm D}, \alpha }
 = \sqrt{(2k_{\rm B}T_{\rm K})/(m_{\rm H}c^2)}{\nu}_{\alpha}$
is the Doppler width of Lyman-$\alpha$,
and ${\nu}_{\alpha}$ is the central Lyman-$\alpha$ frequency
($2.47 \times 10^{15} \ \rm Hz$).
We determine the color temperature and the spin temperature by solving equations
(\ref{eq:Tspin}) and (\ref{eq:color}) iteratively.

\begin{figure}
\epsfig{file=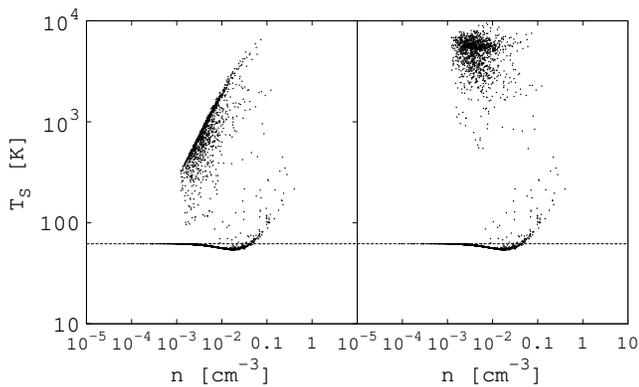, width=8.5cm}
\caption{We plot the 21 cm spin temperature against
gas density. Left: No radiative coupling.
Right:With radiative coupling for $x_{\alpha} = 0.5$.
We use an output at $z=21.7$ when $T_{\rm CMB} =$ 62 K.}
\label{fig:spin}
\end{figure}

We briefly describe the evolution of the \HII region.
Shortly after the central star dies, the kinetic temperature in the \HII region 
is higher than several thousand Kelvin. While high density 
regions within the \HII region recombine and cool quickly,
a large fraction of the \HII region has a low density ($n < 0.01 {\rm cm}^{-3}$),
and thus recombination and gas cooling occur rather slowly.
The gas temperature remains a few thousand Kelvin at 30 Myrs,
and the ionization fraction is down to about 10$\%$, 5$\%$ at 30 Myrs and 50 Myrs 
after the star dies, respectively.

Figure \ref{fig:spin} shows the 21 cm spin temperature against gas
density at a time shortly after the ionizing source has been turned off. 
To see the effect of radiative coupling, we calculate the spin
temperature ignoring (left panel) and accounting for (right panel) radiative coupling.
At low densities, the spin temperatures are close to the CMB temperature for both cases,
because the collisional coupling is weak.
The gas within the \HII regions has been once photo-ionized,
and so the gas kinetic temperature is high, which brings
the spin temperature above 
$\sim$ 100 K for the case without
radiative coupling (left) and
above $\sim 10^3$ K with radiative coupling (right).
Clearly, radiative coupling raises the spin temperatures of
intermediate-density, ionized gases.
We see another branch in the high density, low spin temperature
portion in this phase diagram. It corresponds to the gas outside the relic \HII region,
which has been neutral and hence has a low kinetic temperature.
Note that there are also neutral regions where the spin temperatures are
slightly lower than CMB temperature because of effective collisional coupling.

\begin{figure}
\epsfig{file=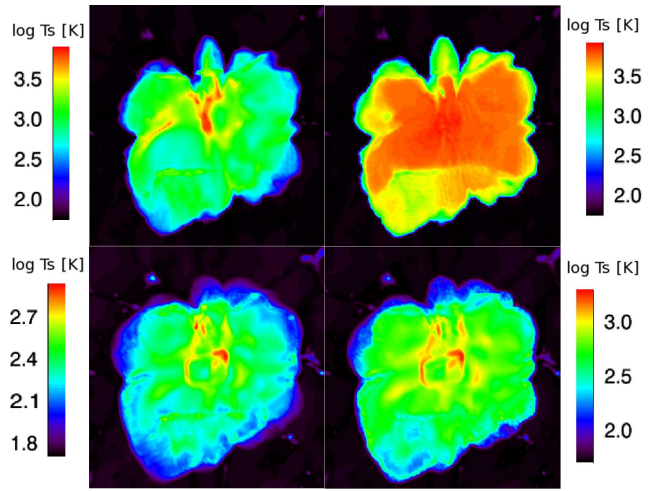, width=8.5cm}
\caption{We show the evolution of the spin temperature.
The output time of the top panels is $z = 21.7$, when the ionizing source 
is switched off,
and that of the bottom panels is $z = 19.5$, about 30 Myrs later.
Left: No radiative coupling.
Right: With radiative coupling ($x_{\alpha} = 0.5$).
The plotted region is 100 $h^{-1}$ comoving kpc on a side.}
\label{map:spin}
\end{figure}

We now observe the simulated \HII region in 21 cm.
The radiative transfer equation for 21 cm radiation for brightness temperature,
$T_{\rm b}$, is given by
\begin{equation}
 T_{\rm b} = T_{\rm CMB} e^{-\tau} + T_{\rm S}(1 - e^{-\tau}).
\label{eq:Tb}
\end{equation}
In practice, we discretize the equation as in Kuhlen et al. (2006)
to calculate the brightness temperature for our simulated \HII region.
The observed flux of 21 cm line can be expressed by the differential brightness 
temperature against the CMB temperature
as $\delta T_{\rm b} = (T_{\rm b} - T_{\rm CMB})/(1+z)$.
To make maps of $T_{\rm S}$ and $\delta T_{\rm b}$,
we use a cubic region of 100$h^{-1}$ comoving kpc on a side, centered at the
\HII region, and then divide the box into $128^3$ cells.
Then the angular resolution of the map is about 0.02 {\bf arcseconds}.
We use the central slice of with a 7.8$h^{-1}$ comoving kpc depth
along the line of sight to make the maps shown in Figure \ref{map:spin}.
{\bf The corresponding spectral width is 190 Hz at $z$=20.} Note that the 
required angular and spectral resolution are out of reach of any planned 
instruments; however, the collective effect of such relic \HII regions might be detectable. 

Figure \ref{map:spin} shows the mass-weighted mean spin temperature at 
two output times, when the ionizing source
died (top panels) and about 30 Myrs after (bottom panels).
The bright, colored area is almost fully ionized (i.e. \HII region)
and the radius of the \HII region is about 50$h^{-1}$ comoving kpc. 
At the top panels, the spin temperature ranges
from a few tens Kelvin ($\sim T_{\rm CMB}$) to several thousand Kelvin.
The top-right panel in Figure \ref{map:spin}
shows that the radiative coupling significantly affects the spin temperature
in the \HII region. After the central star died ($t=0$), the \HII region starts 
recombining, and the gas kinetic temperature decreases. Overall, the mean spin temperature
at $t=30$ Myrs is smaller than at $t=0$.
Hydrodynamic effects are also seen in this high-resolution map. 
Strong shockwaves were driven during the early evolution of the \HII region
(Kitayama et al. 2004; Y07), which are seen as a bright ``ring'' in the bottom
panels in Figure \ref{map:spin}.

\begin{figure}
\epsfig{file=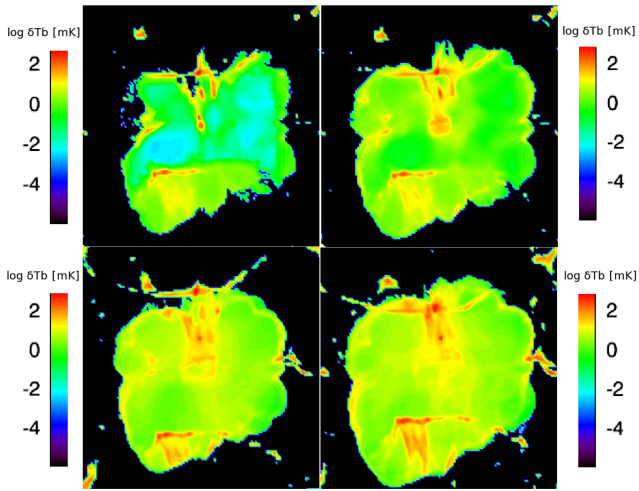, width=8.5cm}
\caption{We plot the time evolution of the 21 cm differential brightness temperature.
  Top-left:Signals at the time of the ionizing source dead ($t=0; z = 21.7$).
  Top-right, bottom-left,and bottom-right are signals at $t=10$ Myrs, 30 Myrs,
  and 50 Myrs.}
\label{map:diff}
\end{figure}

Figure \ref{map:diff} shows the time evolution of the resulting 21 cm signal
for the case with no radiative coupling.
Regions where $\delta T_{\rm b} \leq 0$ are colored in black. 
While a large volume surrounding the \HII region has
$T_{\rm spin} \sim T_{\rm CMB}$ (see Figure \ref{map:spin}), 
which appear in absorption.
We also calculated the differential temperature for the case with efficient 
radiative coupling and find little difference with Figure \ref{map:diff}.
The clear difference in the spin temperature shown in Figure \ref{map:spin}
does not affect the differential temperature.
This is because, as long as $T_{\rm S}$ is much higher than $T_{\rm CMB}$,
the observed brightness is determined by the column density of neutral hydrogen and
the width of 21 cm line, which are independent of $T_{\rm S}$ (e.g., Scott $\&$ Rees 1990).
The large filamentary structure seen in Figure \ref{map:diff} is 
a neutral region which was not completely ionized by the
central star because of its strong self-shielding. 
On average, its spin temperature is not high, hence the
structure is not prominent in Figure \ref{map:spin}, but it shows up
clearly in differential temperature because it has a large 
column density of neutral hydrogen (and hence a large 21 cm 
optical depth).

Figure \ref{map:diff} shows that the relic \HII region is a 21 cm emission source.
In the relic \HII region, $\delta T_{\rm b}$ is as high as several hundred mK,
while in the surrounding neutral regions $\delta T_{\rm b}$ can be as low as $\sim -10$ mK.
The relic \HII region is bright for over 50 Myrs after the source died, 
because the number density of neutral hydrogen increases by recombination
in the relic \HII region, while the spin temperature remains higher than $T_{\rm CMB}$.
The area-averaged $\delta T_{\rm b}$ remains a few tens mK
for over 100 Myrs (at the last output of our simulations).

\section{Discussion}

We have studied the 21 cm signature from relic \HII regions at high redshifts. 
We have computed the spin temperature and 21 cm brightness temperature 
using radiation-hydrodynamics simulations.
We have examined the effect of radiative coupling on the spin temperature 
under simplified assumptions.
We have shown that, whereas the radiative coupling significantly 
affects the spin temperature, it has little effect on 
the 21 cm differential temperature.
The relic \HII region is seen as a bright emission source in 21 cm,
with the differential temperature being up to $\sim 100$ mK for very high angular resolutions.
Previous works estimated the spin temperature and 21 cm signals of \HII regions,
but they did not follow evolution by computing recombination, cooling and the 
expansion of \HII regions dynamically.
We compared our results with those of Nusser (2005), who calculated 
the 21 cm signatures under the isobaric condition.
The overall evolution is similar, but the \HII region in our simulation 
shine in 21 cm for a longer time because recombination occurs slowly
in low density regions.

Early \HII regions are generally very small and thus the individual 21 cm sources 
will not be detected by the Low Frequency Array (LOFAR)\footnote{http://www.lofar.org}
and the Mileura Widefield Array (MWA)\footnote{http://wwww.haystack.mit.edu/ast/arrays/mwa/index.html},
even by the next-generation low-frequency arrays
such as the Square Kilometre Array (SKA)\footnote{http://www.skatelescope.org}.
Minihalos similarly contain little mass, and are not individually detectable. 
A very large, percolated \HII regions or a group of nearby \HII regions 
may be detectable through the effect of strong gravitational lensing (Li et al. 2007).
As we have shown, relic \HII regions are emission sources 
for a long time ($> 50$ Myrs, see Figure \ref{map:spin}), and thus the number density
of such 
21 cm sources at a given frequency can be larger than 
Lyman-$\alpha$ spheres around short-lived stars. 
Therefore, early relic \HII regions could imprint distinct, strong collective 
signatures in the 21 cm background.
In our future work, we will calculate the abundance and the clustering of the relic \HII regions
and derive the 21 cm fluctuation amplitudes.

Finally, we comment that there are some other sources of 21 cm emission and absorption
at high redshifts. Ripamonti et al (2008) study the effect of X-rays from early black holes.
They show that the differential brightness temperature is $\sim 20-30$ mK
at $z \leq 12$. Zaroubi et al (2007) show that X-ray heating produces a
differential brightness temperature of the order of $\sim 5-10$ mK out to a
few comoving Mpc distance from black holes.
These brightness temperatures are smaller, at least locally, than that of relic \HII
regions studied in the present paper.
Thomas \& Zaroubi (2008) consider both primordial black 
holes and Population III stars. They argue that the heating patterns around these 
objects is significantly different.  
A few other exotic models of ionization such as ultra-high energy cosmic rays 
and decaying dark matter particles are proposed (Shchekinov \& Vasiliev 2007). 
Such different sources can likely be distinguished by their different fluctuation spectra.

\section*{acknowledgments}

The simulations were performed at Center
for Computational Cosmology at Nagoya University.
The work is supported in part by the 21st Century
ORIUM Program at Nagoya University, by The Mitsubishi Foundation,
and by the Grants-in-Aid for Young Scientists (A) 17684008, (S) 
20674003 and Grants-in-Aid (C) 17540276 by the Ministry of Education, 
Culture, Science and Technology of Japan.
The work is supported by Grant-in-Aid for
Scientific Research on Priority Areas No. 467 ''Probing
the Dark Energy through an Extremely Wide \& Deep
Survey with Subaru Telescope'' from the Ministry of Education,
Culture, Sports, Science and Technology, Japan.
This work was supported by the Grant-in-Aid for Nagoya
University Global COE Program, "Quest for Fundamental Principles in the
Universe: from Particles to the Solar System and the Cosmos", from the Ministry
of Education, Culture, Sports, Science and Technology of Japan.

\bsp

\label{lastpage}

\end{document}